\documentclass[10pt,letterpaper]{article}
\usepackage{opex3}

\begin{document}

\title{All-Dielectric Optical Nanoantennas}

\author{Alexander~E. Krasnok$^{1,*}$, Andrey~E. Miroshnichenko$^{2,**}$,\\ Pavel~A. Belov$^{1,3}$, Yuri~S. Kivshar$^{1,2}$}
\address{
$^{1}$National Research University of Information Technologies,
Mechanics and Optics, St. Petersburg 197101, Russia\\
$^{2}$Nonlinear Physics Centre, Research School of Physics and
Engineering, Australian National University, Canberra ACT 0200,
Australia\\$^{3}$Queen Mary College, University of London, Mile End
Road, London E1 4NS, United Kingdom}
\email{$^{*}$krasnokfiz@mail.ru, $^{**}$aem124@physics.anu.edu.au}

\begin{abstract}
We study in detail a novel type of optical nanoantennas made of
high-permittivity low-loss dielectric particles. In addition to the
electric resonances, the dielectric particles exhibit very strong
magnetic resonances at the nanoscale, that can be employed in the
Yagi-Uda geometry for creating highly efficient optical
nanoantennas. By comparing plasmonic and dielectric nanoantennas, we
demonstrate that all-dielectric nanoantennas may exhibit better
radiation efficiency also allowing more compact design.
\end{abstract}


%%%%%%%%%%%%%%%%%%%%%%%%%%  body  %%%%%%%%%%%%%%%%%%%%%%%%%%

\section{Introduction}
The recently emerged field of optical nanoantennas is promising for
its potential applications in various areas of nanotechnology. The
ability to redirect propagating radiation and transfer it into
localized subwavelength modes at the
nanoscale~\cite{Hulst_08_OSA,LNov_08_Nature, Koenderink_09_NLetters,
Pakizeh_09_NLetters, Devilez_10_ASCNano, Novotny_10_NatPhot,
Dorfmuller_11_NLetters, Miroshnichenko_11_PSS} makes the optical
nanoantennas highly desirable for many applications. Originally,
antennas were suggested as sources of electromagnetic (EM) radiation
at radio frequencies (RF) and microwaves, emitting radiation via
oscillating currents. Different types of antennas were suggested and
demonstrated for the effective manipulation of the EM
radiation~\cite{Balanis}. Thus, conventional antennas perform a
twofold function as a source and transformation of EM radiation,
resulting in their sizes being comparable with the operational
wavelength. Recent success in the fabrication of nanoscale elements
allows to bring the concept of the RF antennas to optics, leading to
the development of {\em optical nanoantennas} consisting of
subwavelength elements~\cite{Novotny_10_NatPhot}.

\begin{figure} [b]
\centering \centerline{\includegraphics[width=7cm]{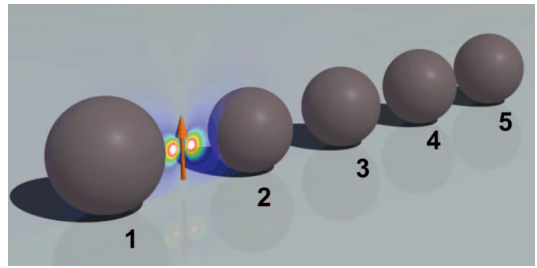}}
\caption{(Color online) Schematic view of an all-dielectric optical
Yagi-Uda nanoantenna, consisting of the reflector (sphere 1) of the
radius $R_r = 75$ nm, and smaller director (spheres 2-5) of the
radii $R_d = 70$ nm. The dipole source is placed equally from the
reflector and the first director surfaces at the distance $D$. The
separation between surfaces of the neighbouring directors is also
equal to $D$. } \label{fig1}
\end{figure}

One of the first fabricated optical nanoantennas was made of
metallic nanorods mimicing the classical analogue of the Yagi-Uda
design~\cite{Hulst_08_OSA,Pakizeh_09_NLetters,Novotny_10_NatPhot,Dorfmuller_11_NLetters}.
It was demonstrated that properly arranged {\em metallic
nanoparticles} can satisfy all the required RF antenna conditions,
and they exhibit high directivity due to the excitation of localized
surface plasmons and strong near-field interaction. Various types of
optical nanoantennas have been discussed in the literature,
including a hybrid design of metallic nanoparticles coupled to
dielectric optical microcavities where high-Q whispering gallery
modes can be used for single-molecule sensors, resonant amplifiers,
nanoconcentrators, energy converters, and dynamical
switches~\cite{Boriskina_11_OPEXPRESS, Boriskina_11_PNAS}. Similar
design, consisting of a dielectric microsphere of $TiO_2$ with
permittivity $\epsilon_1=6.2$ and two silver nanoparticles excited
by a point-like source was also
considered~\cite{Devilez_10_ASCNano}.

In this paper we suggest a novel type of optical nanoantennas made
of all-dielectric elements. We argue that dielectric nanoantennas
can be considered as the best alternative to their metallic
counterparts. First, dielectric materials exhibit {\em low loss} at
the optical frequencies. Second, as was suggested
earlier~\cite{Evlyukhin_10_PRB}, nanoparticles made of
high-permittivity dielectrics may support both {\em electric} and
{\em magnetic} resonant modes. This feature may greatly expand the
applicability of optical nanoantennas for, e.g. detection of
magnetic dipole transitions of molecules. Moreover, it is possible
to realize optical Huygens source \cite{Krasnok_11} consisting of a
point-like electric dipole operating at the magnetic resonance of a
dielectric nanosphere. Such a structure exhibits high directivity
with vanishing backward scattering and polarization independence,
being attractive for efficient and compact designs of optical
nanoantennas.

\section{Design principles for all-dielectric nanoantennas}
We start our analysis by considering a radiation pattern of two
ideal coupled electric and magnetic dipoles. A single point-like
dipole source generates the electric far-field of the following
form~$\mathbf{E}_p =
k^2/(4\pi\epsilon_0r)\exp(ikr)[\vec{\mathbf{p}}-\vec{\mathbf{n}}(\vec{\mathbf{n}}\cdot\vec{\mathbf{p}})]$,
where $\vec{\mathbf{p}}$ is the electric dipole, $k = \omega/c$ is
the wavevector, $\vec{\mathbf{n}}$ is the scattered direction, and
$r$ is the distance from the dipole source. The radiation pattern
$\sigma=\lim\limits_{r\rightarrow\infty}4\pi r^2|E_p|^2$ in the
plane of the dipole $\vec{\mathbf{n}}\times\vec{\mathbf{p}}=0$ is
proportional to the standard figure-eight profile,
$\sigma_{||}\propto|\cos\theta|^2$, where $\theta$ is the scattered
angle. In the plane orthogonal to the dipole
($\vec{\mathbf{n}}\cdot\vec{\mathbf{p}}=0$) the radiation pattern
remains constant and angle independent, $\sigma_{\bot}\propto
const$. Thus, the total radiation pattern of a single dipole emitter
is a torus which radiates equally in the opposite directions. If we
now place, in addition to the electric dipole, an orthogonal
magnetic dipole located at the same point, the situation changes
dramatically. The magnetic dipole $\vec{\mathbf{m}}$ generates the
electric far-field of the form $\mathbf{E}_m =
-(\mu_0/\epsilon_0)^{1/2}k^2/(4r\pi)\exp(ikr)(\vec{\mathbf{n}}\times\vec{\mathbf{m}})$.
Thus, the total electric field is a sum of {\em two contributions}
from both electric and magnetic dipoles $\mathbf{E}_{\rm total} =
\mathbf{E}_p+\mathbf{E}_m$. By assuming that the magnetic dipole is
related to the electric dipole via the relation
$|\vec{\mathbf{m}}|=|\vec{\mathbf{p}}|/(\mu_0\epsilon_0)^{1/2}$,
which corresponds to an infinitesimally small wavefront of a plane
wave often called a Huygens source~\cite{Balanis}, the radiation
pattern becomes $\sigma^H\propto|1+\cos\theta|^2$. This radiation
pattern is quite different compared to that of a single electric
dipole. It is {\em highly asymmetric} with the total suppression of
the radiation in a particular direction, $\theta=\pi$
[$\sigma^H(\pi)=0$], and a strong enhancement in the opposite
direction, $\theta=0$. The complete three-dimensional radiation
pattern resembles a cardioid or apple-like shape, which is also
azimuthally  independent. Such a radiation pattern of the Huygens
source is potentially very useful for various nanoantenna
applications. However, while electric dipole sources are widely used
in optics, magnetic dipoles are less common.

\begin{figure}
\centering \centerline{\includegraphics[width=10.9cm]{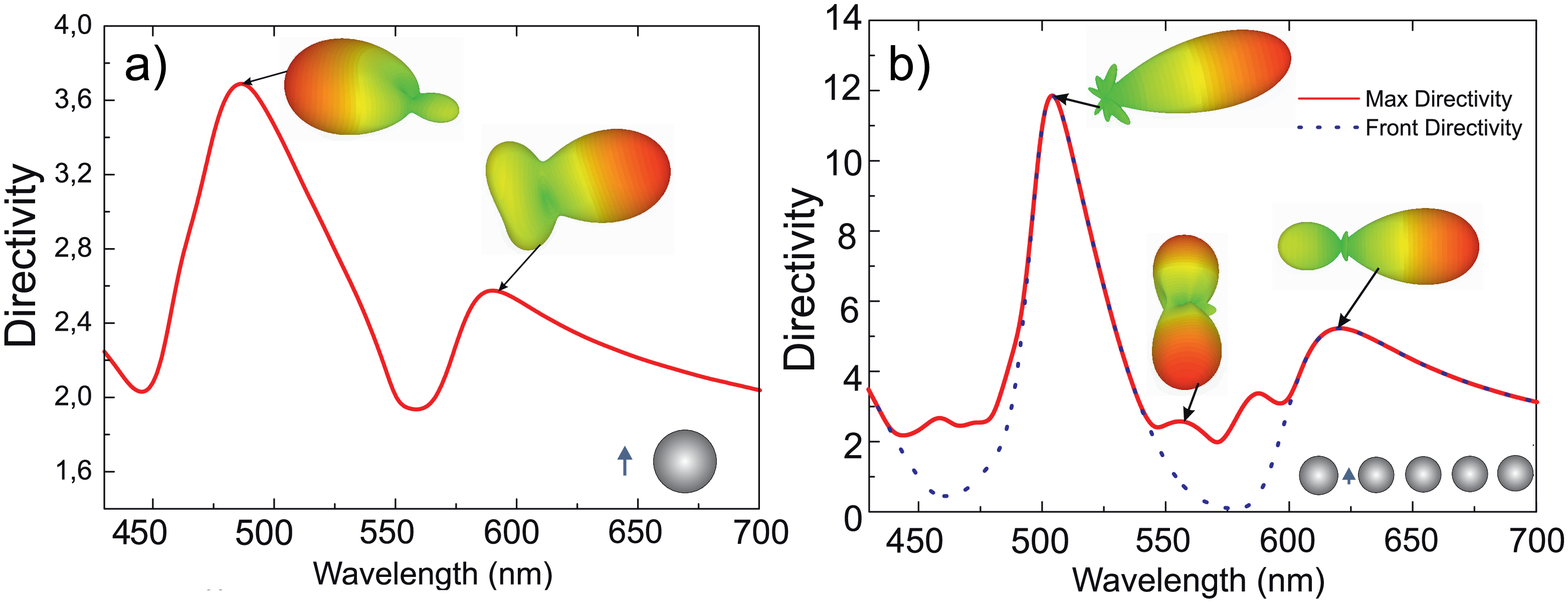}}
\caption{(Color online) Wavelength dependence of the directivity of
two types of all-dielectric  nanoantennas consisting of  (a) single
dielectric nanoparticle, and (b) Yagi-Uda  like design  for the
separation distance $D=70$ nm. Insert shows 3D radiation pattern
diagrams at particular wavelengths.} \label{fig4}
\end{figure}

First, we consider an electric dipole source placed in a close
proximity to a dielectric sphere [see Fig.~\ref{fig1}(a)]. According
to the Mie theory \cite{Bohren}, it can be analytically shown that
high permittivity dielectric nanoparticles exhibit strong magnetic
resonance in the visible range when the wavelength inside the
nanoparticle equals its diameter
$\lambda/n_s\approx2R_s$~\cite{afm_nano}, where $n_S$ and $R_s$ are
refractive index and radius of the nanoparticle, respectively. There
are many dielectric materials with high enough real part of the
permittivity and very low imaginary part, indicating low dissipative
losses. To name just a few, silicon (Si, $\epsilon_{1}=16$),
germanium (Ge, $\epsilon_{1}=20$), aluminum antimonide (AlSb,
$\epsilon_{1}=12$), aluminum arsenide (AlAs, $\epsilon_{1}=10$), and
other.

In our study we concentrate on the nanoparticles made of silicon,
since it satisfies all the requirements being widely used in optics.
The real part of the permittivity of the silicon is about $16$
\cite{Palik}, while the imaginary part is up to two orders of
magnitude smaller than that of nobel metals (silver and gold).
Silicon nanoparticles support strong magnetic resonance in the
visible range for the radius varying from $40$ nm to $80$
nm~\cite{Evlyukhin_10_PRB}.

For such a small radius compared to the wavelength $R_s\ll\lambda$,
the radiation pattern of the silicon nanoparticle in the far field
at the magnetic or electric resonances will resemble that of
magnetic or electric point-like dipole, respectively. Moreover, it
is even possible to introduce magnetic $\alpha^{m}$ and electric
$\alpha^{e}$ polarisabilities~\cite{Evlyukhin_10_PRB,Bohren,
Merchiers_07_PRA} based on  the Mie dipole scattering coefficients
$b_1$ and $a_1$:~$\alpha^{e}=6\pi a_{1} i/k^{3}, \;\alpha^{m}=6\pi
b_{1} i/k^{3}$. Thus,  the dielectric nanoparticle excited by the
electric dipole source at the magnetic resonance  may result in the
total far field radiation pattern which is similar to that of the
Huygens source. Similar radiation patterns can be achieve in light
scattering by a magnetic particle when permeability equals
permittivity $\mu=\epsilon$~\cite{Kerker}. Our result suggests that
even a dielectric nonmagnetic nanoparticle can support two induced
dipoles of equal strength resulting in suppression of the radiation
in the backward direction. Thus, it can be considered as the
simplest and efficient optical nanoantenna with very good
directivity.

In general, both polarisabilities $\alpha^{m}$ and $\alpha^{e}$ are
nonzero in the optical region \cite{Evlyukhin_10_PRB}.  It is known
that for a dipole radiation in the far field the electric and
magnetic components should oscillate in phase to have nonzero energy
flow. In the near field the electric and magnetic components
oscillate with $\pi/2$ phase difference, thus, the averaged Poynting
vector vanishes, and a part of energy is stored in the vicinity of
the source. In the intermediate region, the phase between two
components varies form $\pi/2$ to 0. By placing a nanoparticle close
to the dipole source will change the phase difference between two
components, and, thus, affect the amount of radiation form the near
field. In the case of plasmonic nanoparticles which exhibit electric
polarizability only, there is an abrupt phase change from 0 to $\pi$
in the vicinity of the localized surface plasmon resonance, which
makes it difficult to tune  plasmonic nanoantennas for optimal
performance. The dependence of the scattering diagram on the
distance between the electric dipole source and metallic
nanoparticle was studied in Ref.~ \cite{Rolly_11_OptLett}. On
contrary, in the case of nanoparticles with both electric and
magnetic polarisabilities, it is possible to achieve more efficient
radiation from the near to far field zone, due to subtle phase
manipulation. {\em This is exactly the case of the dielectric
nanoparticles}.

Any antenna is characterized by two specific properties, Directivity
$(D)$ and Radiation Efficiency $(\varepsilon_{rad})$, defined as
\cite{Novotny_10_NatPhot, Balanis}
\begin{equation}
D=\frac{4\pi}{P_{\mbox{rad}}}\mbox{Max}[p(\theta,\varphi)], \;\;
\eta_{rad}=\frac{P_{\mbox{rad}}}{P_{\mbox{rad}}+P_{\mbox{loss}}},
\end{equation}
where $P_{\mbox{rad}}$ and $P_{\mbox{loss}}$ are integrated radiated
and absorbed powers, respectively, $\theta$ and $\varphi$ are
spherical angles, and $p(\theta,\varphi)$ is the radiated power in
the given direction  $\theta$ and/or $\varphi$.

\begin{figure}
\centering \centerline{\includegraphics[width=10.8cm]{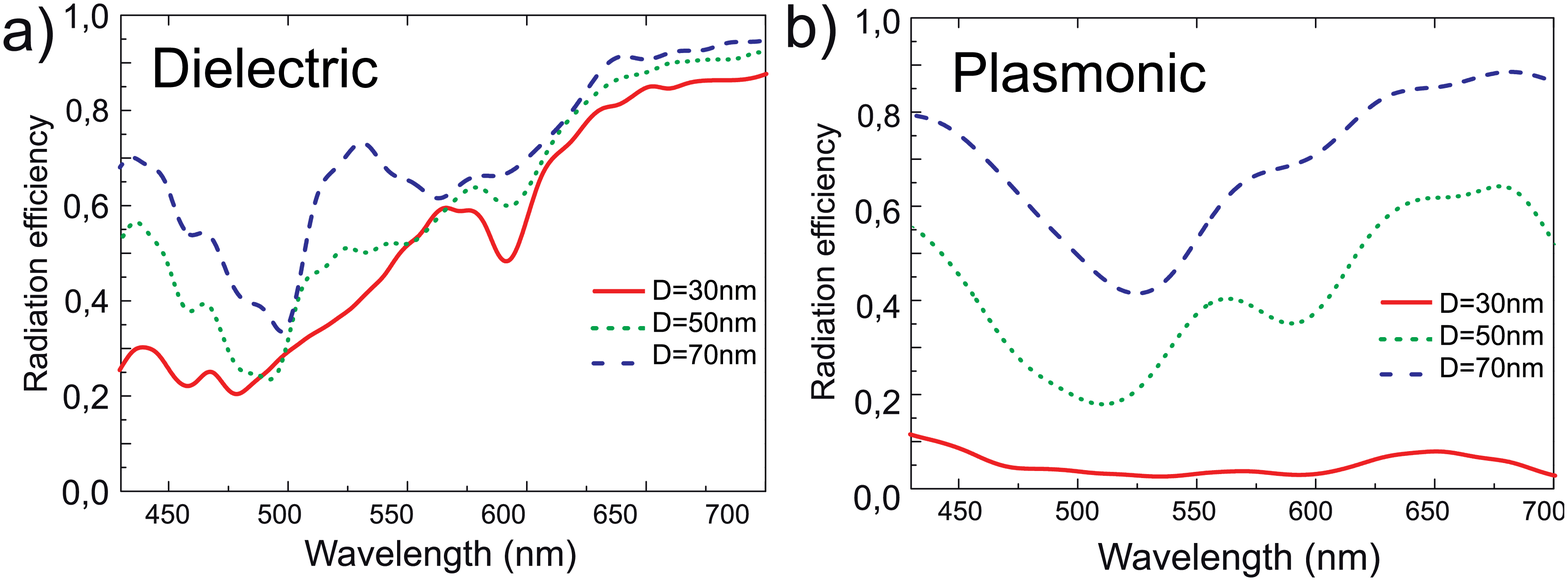}}
\caption{(Color online) Radiation efficiencies of (a) dielectric
(Si) and (b) plasmonic (Ag) Yagi-Uda optical nanoantennas of the
same geometrical designs for various values of the separation
distance $D$.} \label{fig5}
\end{figure}

The Directivity measures the power density of the antenna radiated
in the direction of its strongest emission, while Radiation
Efficiency measures the electrical losses that occur throughout the
antenna at a given wavelength. To calculate these quantities
numerically for the structures shown in Fig.~\ref{fig1}, we employ
the CST Microwave Studio. To get reliable results, we model the
electric dipole source by a Discrete Port coupled to two PEC
nanoparticles.

In Fig. ~\ref{fig4}(a) we show the dependence of the Directivity on
wavelength for a single dielectric nanoparticle excited by a
electric dipole source. Two inserts demonstrate 3D angular
distribution of the radiated pattern $p(\theta,\varphi)$
corresponding to the local maxima. In this case, the system radiates
predominantly to the forward direction at $\lambda=590$ nm, while in
another case, the radiation is predominantly in the backward
direction at $\lambda=480$ nm. In this case, the total electric
dipole moment of the sphere and point-like source and the magnetic
dipole moment of the sphere oscillate with the phase difference
$\mbox{arg}(\alpha^{m})-\mbox{arg}(\alpha^{e})=1.3\mbox{rad}$,
resulting in the destructive interference in the forward direction.
At the wavelength $\lambda = 590$ nm the total electric and magnetic
dipole moments oscillate in phase and produce Huygens-source-like
radiation pattern with the main lobe directed in the forward
direction.

By adding more elements to the silicon nanoparticle, we can enhance
the performance of all-dielectric nanoantennas. In particular, we
consider a dielectric analogue of the Yagi-Uda design (see
Fig.\ref{fig1}) consisting of four directors and one reflector. The
radii of the directors and the reflector are chosen to achieve the
maximal constructive interference in the forward direction along the
array. The optimal performance of the Yagi-Uda nanoantenna should be
expected when the radii of the directors correspond to the magnetic
resonance, and the radius of the reflector correspond to the
electric resonance at a given frequency, with the coupling between
the elements taken into account. Our particular design consists of
the directors with radii $R_d = 70$ nm and the reflector with the
radius $R_r = 75$ nm. In Fig.~\ref{fig4}(b) we plot the directivity
of all-dielectric Yagi-Uda nanoantenna vs. wavelength with the
separation distance $D = 70$ nm. Inserts demonstrate the 3D
radiation patterns at particular wavelengths. We achieve a strong
maximum at $\lambda = 500$ nm. The main lobe is extremely narrow
with the beam-width about $40^\circ$ and negligible backscattering.
The maximum does not correspond exactly to either magnetic or
electric resonances of a single dielectric sphere, which implies the
importance of the interaction between constitutive nanoparticles.

\begin{figure}
\centering \centerline{\includegraphics[width=10.8cm]{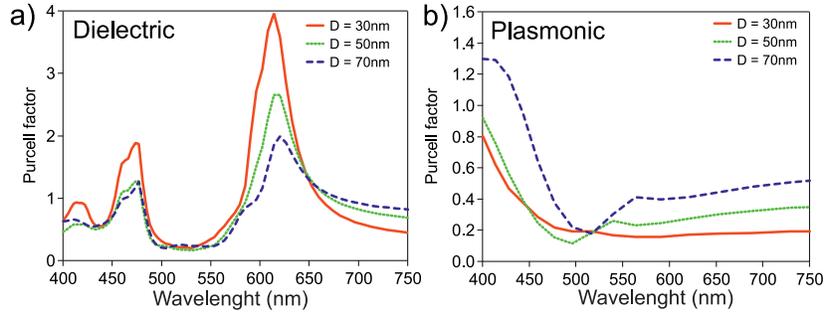}}
\caption{(Color online) Purcell factor of all-dielectric Yagi-Uda
nanoantenna vs wavelength for various values of the separation
distance $D$.} \label{fig6}
\end{figure}

As the next step, we study the performance of the all-dielectric
nanoantennas for different separation distances $D$, and compare it
with {\em a plasmonic analogue} of the similar geometric design made
of silver nanoparticles. According to the results summarized in
Fig.~\ref{fig5}, the radiation efficiencies of both types of
nanoantennas {\em are nearly the same} for larger separation of
directors $D=70$ nm with the averaged value $70\%$. Although
dissipation losses of silicon are much smaller than those of silver,
the dielectric particle absorbs the EM energy by the whole spherical
volume, while the metallic particles absorb mostly at the surface.
As a result, there is no big difference in the overall performance
of these two types of nanoantennas for relatively large distances
between the elements. However, the difference becomes {\em very
strong} for smaller separations. The radiation efficiency of the
all-dielectric nanoantenna is insensitive to the separation distance
[see Fig.~\ref{fig5} (a)]. On contrary, the radiation efficiency
drops significantly for metallic nanoantennas [see Fig.~\ref{fig5}
(b)].

Finally, we investigate the modification of the transition rate of a
quantum point-like source placed in the vicinity of dielectric
particles. For electric-dipole transitions and in the weak-coupling
regime, the normalised spontaneous decay rate $\Gamma/\Gamma_0$,
also known as Purcell factor, can be calculated classically as the
ratio of energy dissipation rates of an electric dipole
$P/P_0$~\cite{Novotny}. Here, $\Gamma_0$ and $P_0$ correspond to
transition rate of the quantum emitter and energy dissipation rate
of the electric dipole in free space~\cite{Chew}. In the limit of
the intrinsic quantum yield of the emitter close to unity, both
ratios become equal to each other  $\Gamma/\Gamma_0=P/P_0$, which
allows us to calculate the Purcell factor in the classical
regime~\cite{Novotny}.  We have calculated the Purcell factor by
using both, numerical and analytical approaches. Numerically, by
using the CST Microwave Studio we calculate the total radiated in
the far-field and dissipated into the particles powers and take the
ratio of their sum to the total power radiated by the electric
dipole in free space. Analytically, we employed the generalised
multiparticle Mie solution~\cite{Xu} adapted for the electric dipole
excitation~\cite{Dulkeith}. We verified that both approaches produce
similar results. In Fig.~\ref{fig6} we show calculated Purcell
factor of the all-dielectric Yagi-Uda nanoantenna vs. wavelength for
various separation distances. We observe that, by decreasing the
separation between the directors, the Purcell factor becomes
stronger near the magnetic dipole resonance. Note here, that a
plasmonic analogue of the same nanoantenna made of Ag exhibits low
Purcell factor less than one, see, for example,
Ref.~\cite{St_11_OSA}. Thus, such relatively high Purcell fcan be
employed for efficient photon extraction from molecules placed near
all-dielectric optical nanoantennas.

\section{Conclusion}
We have suggested a novel type of optical nanoantennas made of
dielectric nanoparticles. Such all-dielectric nanoantennas
demonstrate a number of key advantages over their metallic
counterparts, including much lower dissipation losses and strong
optically-induced  magnetization. We have analyzed an all-dielectric
analogue of the plasmonic Yagi-Uda nanoantenna consisting of an
array of nanoelements, and have demonstrated very high directivity
with a smaller number of directors. Moreover, lower dissipation
losses and localization of the electromagnetic field inside the
nanoparticles allows to reduce the distance between the adjacent
elements even further, without compromising the performance.

\section{Acknowledgments} The authors thank C.R. Simovski for useful
discussions, and acknowledge a support from the Ministry of
Education and Science in Russia and the Australian Research Council.

\end{document}